\documentclass[aps,pre,twocolumn,superscriptaddress,showpacs,nolongbibliography]{revtex4-2}
\usepackage{amsmath}
\usepackage{amssymb}
\usepackage{MnSymbol}
\usepackage{graphicx}
\usepackage{dcolumn}
\usepackage{bm}
\usepackage[normalem]{ulem}

\usepackage[colorlinks, linkcolor = blue, citecolor = blue, filecolor = blue, urlcolor = blue]{hyperref}

\usepackage{orcidlink}

\begin{document}

\title{Systematic analysis of critical exponents in continuous dynamical phase transitions of weak noise theories}

\author{Timo Schorlepp \orcidlink{0000-0002-9143-8854}}
\email{timo.schorlepp@nyu.edu}
\affiliation{Courant Institute of Mathematical Sciences, New York University, 251 Mercer St, New York, NY 10012, USA}

\author{Ohad Shpielberg \orcidlink{0000-0002-7911-4830}}
\email{ohads@sci.haifa.ac.il}
\affiliation{Department of Mathematics and Physics, University of Haifa at Oranim, Kiryat Tivon 3600600, Israel}
\affiliation{Haifa Research Center for Theoretical Physics and Astrophysics,
University of Haifa, Abba Khoushy Ave 199, Haifa 3498838, Israel}

\date{\today}

\begin{abstract}
Dynamical phase transitions are nonequilibrium counterparts of thermodynamic phase transitions and share many similarities with their equilibrium analogs. In continuous phase transitions, critical exponents play a key role in characterizing the physics near criticality. This study aims to systematically analyze the set of possible critical exponents in weak noise statistical field theories in $1+1$ dimensions, focusing on cases with a single fluctuating field. To achieve this, we develop and apply the Gaussian fluctuation method, avoiding reliance on constructing a Landau theory based on system symmetries. Our analysis reveals that the critical exponents can be categorized into a limited set of distinct cases, suggesting a constrained universality in weak noise-induced dynamical phase transitions. We illustrate our findings in two examples: short-time large deviations of the Kardar--Parisi--Zhang equation, and the weakly asymmetric exclusion process on a ring within the framework of the macroscopic fluctuation theory.
\end{abstract}

\maketitle

\section{Introduction}

Over the past few decades, significant progress has been made in the field of nonequilibrium physics, a domain that now stands at the forefront of modern theoretical and experimental research. The study of systems far from equilibrium has led to the development of several robust theoretical frameworks, each addressing different aspects and subclasses of nonequilibrium phenomena. For instance, the theory of stochastic thermodynamics has provided deep insights into the behavior of small systems subject to thermal fluctuations \cite{seifert:2012,van2015ensemble}, while the macroscopic fluctuation theory (MFT) has offered a systematic approach to understanding large deviations in driven diffusive systems \cite{bertini-de-sole-gabrielli-etal:2015,derrida2007non}. Additionally,  hydrodynamic theories have been pivotal in describing the behavior of fluids and gases away from equilibrium \cite{spohn2012large,spohn2014nonlinear,doyon2020lecture}, and the burgeoning field of active matter physics has shed light on the dynamics of systems composed of self-propelled particles \cite{marchetti-joany-ramaswamy-etal:2013,bechinger2016active,nardini2017entropy}. These theories, along with advances in the understanding of nonequilibrium steady states, fluctuation theorems, and dynamical phase transitions (DPTs), have collectively enriched our comprehension of the complex behaviors exhibited by nonequilibrium systems, paving the way for further exploration and discovery.

Despite the significant advancements in nonequilibrium physics, we are still far from possessing a unified theory with a comparably comprehensive scope as equilibrium statistical mechanics. In equilibrium systems, statistical mechanics allows us to calculate thermodynamic variables by counting the number of microstates corresponding to a given macrostate, leading to well-defined quantities such as entropy and free energy. However, the powerful idea of state counting does not generally extend to nonequilibrium systems.
In the 1970s, David Ruelle proposed shifting the focus from states to trajectories by introducing the idea of counting trajectories associated with time-integrated quantities, putting forward large deviation functions as nonequilibrium analogs of thermodynamic potentials  \cite{ruelle2004thermodynamic,touchette:2009}. In what follows, we take advantage of this analogy. 

Equilibrium phase transitions, especially second-order or continuous transitions, are of profound importance in statistical mechanics. 
At these transitions, the thermodynamic potential, such as the free energy, becomes non-analytic at a critical point, signaling the emergence of long-range correlations and fluctuations that govern the system's behavior. Critical exponents (CEs), together with the singular scaling function, capture how physical quantities like the correlation length and susceptibility diverge when approaching the critical point, thus encapsulating the physics of the system close to criticality. 
The conceptual framework of equilibrium phase transitions, with its focus on critical points and scaling exponents, extends naturally to nonequilibrium systems through the Ruelle construction. This approach leads to the identification of DPTs, which mirror thermodynamic phase transitions in their structure and significance. 

The study of thermodynamic phase transitions has been remarkably successful, with a strong agreement between theoretical predictions, numerical simulations, and experimental observations. Diverging quantities such as susceptibilities, specific heat, and correlation length near critical points are well-characterized, allowing for precise verification across different methodologies \cite{ma:2018,kardar:2007,cardy:1996,cardy:2012,stanley:1971}. In contrast, obtaining and verifying theoretical results in the realm of large deviations and DPTs presents a greater challenge. Experimental validation is often difficult due to the rarity of large deviation events, which require extensive data collection and precise control over system parameters. Even numerical approaches, while valuable, can be computationally intensive and may struggle to capture the full complexity of these phenomena. Consequently, bridging the gap between theory and empirical observation in the context of large deviations remains a significant hurdle. 

Some progress has been made in understanding DPTs in Langevin systems with weak noise. A weak noise Langevin system indicates that stochastic fluctuations are small compared to the deterministic forces, allowing the noise to be treated as a perturbation. This simplification makes it easier to obtain analytical and numerical results, as the dynamics can often be captured by focusing on rare but significant fluctuations that dominate the system's behavior. For several important models, different approaches have successfully identified and validated the CEs characterizing these transitions \cite{gerschenfeld2011current,shpielberg-nemoto-caetano:2018,baek2017dynamical,baek2018dynamical}. A prevalent method has been to emulate the Landau theory of phase transitions~\cite{landau:1937}, which relies on symmetry considerations and the identification of an appropriate order parameter \cite{smith-kamenev-meerson:2018,agranov2023tricritical,meibohm-esposito:2022,meibohm-esposito:2023,vadakkayil-esposito-meibohm:2024}. However, the applicability of the Landau theory is limited if the relevant symmetries or the order parameter are difficult to identify. Interestingly, there are instances where different spontaneous symmetry breaking mechanisms have led to the same CEs \cite{shpielberg-nemoto-caetano:2018}, suggesting a greater degree of universality than previously expected. This has led to conjectures that the universality classes of DPTs might be broader and more interconnected than those of traditional thermodynamic phase transitions.

In this work, we develop a systematic approach to extract the CEs of DPTs in weak noise Langevin systems, without relying on the Landau theory or invoking the symmetries of the system. The method provides a general framework for analyzing these transitions, offering a versatile tool that can be applied even when the traditional symmetry-based approach is not feasible. We demonstrate the effectiveness of our approach through examples drawn from the MFT and the short-time behavior of the Kardar--Parisi--Zhang (KPZ) equation, both of which have been previously studied and shown to exhibit DPTs in numerous setups. By applying our method to these established models, we offer new insights into the universality and critical behavior of these systems.

This work is structured as follows: In section~\ref{sec:setup}, we set up the general system under study and present our results in detail. Section~\ref{Sec:Infinite range Ising} serves as a pedagogical introduction to our method, using the Curie-Weiss (CW) model as a simple illustrative example. In section~\ref{Sec:General theory}, we introduce our method for extracting the CEs in weak noise Langevin systems, focusing on a single spatio-temporal field in $1+1$ dimensions. Sections \ref{sec:KPZ} and \ref{sec:MFT}  demonstrate our approach explicitly through examples from the short-time KPZ equation and the MFT, respectively. Finally, in section~\ref{sec:Summary}, we summarize our results and discuss their implications within the broader context of critical behavior in nonequilibrium physics.

\section{Setup and results \label{sec:setup}}

In this study, we consider weak noise Langevin systems, focusing on a single field $\phi = \phi(x,t)$ in $1+1$ dimensions. Treatment of higher-dimensional systems is skipped to maintain clarity in the notation and focus on the core message. The dynamics of the system are governed by the Langevin  equation 
\begin{equation}
\label{eq:Langevin equation}
    \partial_t \phi = F[\phi] + \sqrt{\varepsilon G[\phi]} \,  \xi(x,t) , 
\end{equation}
 where $F[\phi]$ represents the deterministic force acting on the field, $G[\phi]$ characterizes the possibly field-dependent fluctuations, and $\xi (x,t)$
 is standard Gaussian space-time white noise with $\left \langle \xi(x,t) \right \rangle = 0$, $\left \langle \xi(x,t) \xi(x',t')\right \rangle = \delta(x-x') \delta(t-t')$. The functionals $F,G$ may include spatial derivative terms, as well as non-local terms \cite{dandekar2023dynamical}.  
 The parameter $\varepsilon > 0$ quantifies the relative strength of the stochastic noise compared to the deterministic dynamics. The path probability $P[\phi]$ of observing the path $\phi$ through space and time under the Langevin equation~\eqref{eq:Langevin equation} in the It{\^o} interpretation is given by
 \begin{eqnarray}
 \label{eq:path prob h}
      P[\phi] &\asymp& \exp\left\{-\varepsilon^{-1} \, S[\phi] \right\}.
  \end{eqnarray}
To obtain the explicit form of the action functional $S[\phi]$, one may adopt the It{\^o} discretization scheme and employ the Martin--Siggia--Rose formalism~\cite{martin-siggia-rose:1973}, yielding the classical Onsager--Machlup~\cite{machlup-onsager:1953} or Freidlin--Wentzell~\cite{freidlin-wentzell:2012} action
\begin{align}
    S[\phi]= \frac{1}{2}\int \mathrm{d} x \int \mathrm{d} t \left(\partial_t \phi - F[\phi] \right) G^{-1}[\phi] \left(\partial_t \phi - F[\phi] \right)\,.
    \label{eq:action-func}
\end{align}
In this work we consider $\varepsilon \ll 1$, where the discretization scheme becomes irrelevant to leading order in $\varepsilon$, cf.~\cite{langouche-roekaerts-tirapegui:1982,itami-sasa:2017,cugliandolo-lecomte:2017,gladrow-keyser-adhikari-etal:2021} for detailed discussions of the impact of different stochastic integral conventions and path integral discretization schemes. Hence, crucially, the discretization choice is immaterial to both the form of the action $S[\phi]$ and to the CEs in the weak noise limit $\varepsilon \ll 1$ studied here. 
   
 Our objective is to investigate the physics of all system trajectories that obey the condition 
 \begin{equation}
 \label{eq:z-condition-gen}
 \int \mathrm{d} x \int \mathrm{d} t \; Z[\phi] = z,    
 \end{equation}
  where $Z[\phi]$ is a functional of the field $\phi$, for given values of $z$. To this end, we set out to determine $\mathcal{P}(z)$ -- the probability (density) of observing a trajectory that satisfies the constraint~\eqref{eq:z-condition-gen}. Formally, $\mathcal{P}(z)$ is the conditional sum over all the path probabilities~\eqref{eq:path prob h} that satisfy~\eqref{eq:z-condition-gen}. This is essentially the Ruelle construction of counting trajectories. In general, finding the expression of $\mathcal{P}(z)$ is a challenging task. We remark at this point that the constraint~\eqref{eq:z-condition-gen} could for instance depend on the full field configuration in space and time, or only on the field configuration in space at a time $T$, or only on the field evaluated at a single point in space and time, and so forth. We will see below that distinguishing these cases is important for the proper determination of the CEs.

 In the weak noise limit, where $\varepsilon \ll 1 $, the conditional sum of $\mathcal{P}(z)$ is typically dominated by a single fluctuation known as the optimal fluctuation or a soliton solution. In this case, one expects a large deviation principle 
  \begin{eqnarray}
       \mathcal{P} (z)  &\asymp& \exp \left\{-
\varepsilon^{-1} I(z) \right\},
  \end{eqnarray}
  where $I(z)$ is the large deviation rate function, and it is dominated by the optimal fluctuation $\bar{\phi}$ -- the system trajectory that minimizes the action $S[\phi]$ subject to the constraint~\eqref{eq:z-condition-gen}. In other words, $I(z) = S[\bar{\phi}] + O(\varepsilon)$.  The optimal fluctuation method in the weak noise limit constitutes a dramatic simplification in the analytical determination of $\mathcal{P}(z)$. 
  Although calculating $\mathcal{P}(z)$ in this limit may remain technically challenging, the optimal fluctuation method has proven to be a powerful method in numerous cases \cite{bodineau2004current,bettelheim2022inverse,mallick2022exact,zarfaty2016statistics}. 
  
Indeed, as $\varepsilon \to 0$, the large deviation function $I(z)$ is dominated by the optimal fluctuation associated with the condition~\eqref{eq:z-condition-gen}, discarding the conditional sum altogether. Typically, when taking into account ``finite size'' corrections in $\varepsilon$, i.e.\ taking the conditional sum, one obtains corrections of order $O(\varepsilon)$ to the large deviation function.  
The state of affairs changes when $I_0(z)\equiv S[\bar{\phi}]$ becomes non-analytic at a critical value $z = z_{\mathrm{c}}$. Such a point constitutes a DPT and signifies a dramatic change in the  properties of the optimal fluctuation, in turn leading to a dramatic change in the physical properties of the conditioned system. If the DPT is continuous, then as $\varepsilon \downarrow 0$ and $z \uparrow z_{\mathrm{c}}$ (where $z < z_{\text{c}}$ is assumed without loss of generality), one may expect an anomalous correction to the saddle solution 
\begin{align}
\begin{gathered}
\label{eq:I singular scaling form}
   I(z)  = I_0(z) + \varepsilon^\lambda I_{\text{s}} (\delta z / \varepsilon^\theta )
    \\ 
    \; + \; \textrm{subleading contributions}.
\end{gathered}
\end{align}
Here, $\delta z = z_{\mathrm{c}} - z > 0$,  $I_{\text{s}}$ is the singular scaling function of the large deviation function, and $\lambda,\theta>0$ are the CEs. Note that the singular scaling function is expected to satisfy $\varepsilon^\lambda I_\text{s}(\delta z / \varepsilon^\theta)\propto \varepsilon$ for $\delta z$ finite and $\varepsilon\downarrow 0$, which determines the scaling of $I_{\text{s}}(\mu \gg 1)$.  The main goal of this work is providing a recipe to determine the CEs $\lambda,\theta$. It should be stressed that $\lambda,\theta$ are not the textbook CEs $\alpha,\beta,\gamma,\delta,\nu$~\cite{ma:2018,kardar:2007,cardy:1996,cardy:2012,stanley:1971} since we are interested in the rate function $I$ itself, i.e.\ the analog of the free energy density, instead of its order parameter associated derivatives. We postpone clarifying the relations between the CEs $\lambda,\theta$ and the textbook CEs to the summary in section~\ref{sec:Summary}.   

Section~\ref{Sec:General theory} provides a detailed recipe to extracting the CEs $\lambda, \theta$, the results of which are summarized in table~\ref{tab:ce-results} in the simplest case. It stands on two main legs, taking a page from the theory of second order thermodynamic phase transitions. The first leg assumes the Ginzburg criterion \cite{kardar:2007,cardy:1996}, stating that\, for a second-order DPT, the dominance of the fluctuations becomes comparable to the mean field analysis in the second derivative of the large deviation function in~\eqref{eq:I singular scaling form}. The Ginzburg criterion implies the relation $\lambda = 2\theta$ for a second-order DPT. The second leg recalls Le Chatelier's principle \cite{callen1991thermodynamics,shpielberg2016chatelier}, or simply put, the stability analysis of the optimal fluctuation. As~$\delta z \downarrow  0$, stability of the optimal fluctuation~$\bar{\phi}$ as a local minimum is compromised by a set of subleading fluctuations. Accounting for the contribution of the destabilizing fluctuations allows to acquire the scaling form of~$I_\text{s}$ when $\delta z \gg \varepsilon^\theta$,
and in this limit the singular part of~\eqref{eq:I singular scaling form} becomes linear in $\varepsilon$. This step leads to a second relation between the CEs $\lambda , \theta$. Therefore, the two legs allow to find the CEs $\lambda,\theta$ explicitly. We name this procedure the Gaussian fluctuation method. We compare its range of applicability to the traditional Landau theory in table~\ref{tab:comparison}.

\begin{table}
\caption{
Summary of the calculated critical exponents~$\lambda, \theta$ in~\eqref{eq:I singular scaling form} for the weak-noise Langevin equation~\eqref{eq:Langevin equation} restricted by the condition \eqref{eq:z-condition-gen},
on a domain of size $L$ and in time $T$, as derived in section~\ref{Sec:General theory}, in the simplest case of a second-order continuous DPT and $n=1$ in the Taylor expansion~\eqref{eq: k0 omega0 f} of the unstable mode coefficient $g_z$. Case A also includes ``$L$ infinite, $T$ finite'' as well.} 
\label{tab:ce-results}
\begin{ruledtabular}
\begin{tabular}{c|ccc}
case & characterization & $\lambda$ & $\theta$\\[.5pt]
\hline\\[-5pt]
A& $L$ finite, $T$ infinite & $4/3$ & $2/3$\\[2pt]
B & $L$ and $T$ finite & $1$ & $1 / 2$\\[2pt]
C & $L$ and $T$ infinite & $2$ & $1$
\end{tabular}
\end{ruledtabular}
\end{table}

Importantly, this study shows that the values of the CEs depends on the Fourier decomposition of the fluctuations around the optimal fluctuation $\delta \phi(x,t) = \phi(x,t)-\bar{\phi}(x,t)$. More precisely, the universality class of the CEs $\lambda,\theta$ depends on counting the effective number $d_\perp$ of continuous Fourier frequencies in $\delta \phi(x,t)$, which in turn depends on the setup~\eqref{eq:Langevin equation} and path condition~\eqref{eq:z-condition-gen}. This is derived in section~\ref{Sec:General theory}, and demonstrated in both section~\ref{sec:MFT} and section~\ref{sec:KPZ}. 

Before we tackle weak-noise Langevin systems with the described method, we start with a pedagogical example of the CW model at thermal equilibrium. 

\begin{table*}
    \renewcommand{\arraystretch}{1.3}
    \centering
    \begin{tabular}{p{3cm}|p{6.5cm}|p{6.5cm}}
        \hline\hline
        & \textbf{Gaussian Fluctuation Method} & \textbf{Landau Theory} \\
        \hline
        \textbf{Conditions to use} & Knowledge of transition order \( G_{\text{c}} \geq 2 \), and assuming the unstable modes in the Gaussian terms vanish linearly in  $ \delta z$ ($n=1$)
        &
        Identification of the order parameter, symmetries, and exact  mean-field solution \( \bar{\phi}(x,t) \) \\
        \hline
        \textbf{Predicts} & Critical exponents not directly related to the order parameter: \( \theta, \lambda, \alpha, \nu \) & All critical exponents: \( \theta, \lambda, \alpha, \beta, \gamma, \delta, \nu \) \\
        \hline\hline
    \end{tabular}
    \caption{Comparison between Landau theory and the Gaussian fluctuation method presented here. The Landau theory provides a complete prediction of all critical exponents but requires an explicit mean-field solution, which is often challenging to obtain. The Gaussian fluctuation method, while limited in scope, is more broadly applicable, especially when the dynamical phase transition is complex and lacks an analytically tractable mean-field description.}
    \label{tab:comparison}
\end{table*}

\section{The Curie-Weiss model 
\label{Sec:Infinite range Ising}}

Extracting the CEs of the weak noise Langevin equation~\eqref{eq:Langevin equation} has been achieved in different examples in the literature. Going beyond these specific examples remains technically challenging, calling for a broader approach that rises above technical difficulties.  Before addressing this main goal, it is useful to consider the CW model as a pedagogical example, where technical details are kept to a minimum. 

The CW model was proposed as a toy mean-field model exhibiting a ferromagnetic phase transition~\cite{kochmanski2013curie}. Thus, it serves as perhaps the simplest model to study phase transitions and spontaneous symmetry breaking. The mean-field approximation will allow to extract the CEs of the CW model, and pave the way for extracting CEs in the weak-noise Langevin equation~\eqref{eq:Langevin equation}. 

Consider $N$ interacting classical spins $s_i$ that take the values $s_i=\pm 1$ for $i=1,...,N$. The equilibrium properties of the model are described by the Hamiltonian
\begin{equation}
\label{eq:C-W model Ham}
\mathcal{H}\left(\left\{s_i\right\} \right) = - \frac{J}{2N} S^2\,, \quad S = \sum^N_{i=1} s_i\,.
\end{equation}
Here $J>0$, which induces a ferromagnetic order, describes the strength of interactions between the spins.
It is useful to define the dimensionless parameter $z= J/k_B T > 0$, where $T$ is the system's temperature and  $k_B$ is the Boltzmann constant. The notation $z$ is chosen to stress the analogy to the constraint~\eqref{eq:z-condition-gen} in the weak-noise Langevin setup of the previous section, whereas $N$ here corresponds to $1 /\varepsilon$ in the weak-noise setting~\eqref{eq:Langevin equation}. Through~\eqref{eq:C-W model Ham}, one can express the partition function ${\cal Z}_N$ of the CW model as
\begin{eqnarray}
    {\cal Z}_N (z) &=& 
    \sum_{ \left\{s_i \right\} \in  \left\{\pm 1 \right\}^N} e^{-\mathcal{H}\left(\left\{s_i\right\} \right) / (k_B T)} =
    \sum_{ \left\{s_i \right\} \in  \left\{\pm 1 \right\}^N} e^{\frac{1}{2} \Sigma^2 }  \nonumber 
    \\
    &= &  \frac{1}{\sqrt{2\pi}} \int_{-\infty}^\infty \mathrm{d} x  \sum_{ \left\{s_i \right\} \in  \left\{\pm 1 \right\}^N}  e^{-\frac{1}{2}(x+\Sigma)^2 +\frac{1}{2}\Sigma^2} \,.
    \label{eq:HS transform}
\end{eqnarray}
Here $\Sigma = \sqrt{z/N}S$, and in the second line of~\eqref{eq:HS transform} we have used a Hubbard--Stratonovich transform to avoid summing over the quadratic terms $s_i s_j$ in the argument of the exponent, i.e. the $\Sigma^2$ term. Changing the integration variable to $m=x/ \sqrt{N z} $ we find
\begin{eqnarray}
    {\cal Z}_N(z) = \sqrt{\frac{N z}{2\pi}} \int_{-\infty}^\infty \mathrm{d} m \; e^{-N g_z(m)}\,,
    \\  \nonumber 
    g_z(m) = \frac{1}{2} z m^2 - \log \left( 2 \cosh (zm)\right)\,.
    \label{eq:g-def-cw-model}
\end{eqnarray}
In the large $N$ limit, this expression suggests that the partition function ${\cal Z}_N(z)$ is dominated by the minimum of $g_z(m)$. Therefore, the free energy per spin divided by the temperature $\mathfrak{f}_N(z)$ is given by 
\begin{eqnarray}
   \mathfrak{f}_N(z) = F_N(z) / N k_B T &=& -\frac{1}{N} \log {\cal Z}_N(z) 
    \\ \nonumber
    &\simeq& g_z(\bar{m})  + \textrm{subleading terms},  
\end{eqnarray}
where $\bar{m} =\bar{m}(z)$ indicates the global minimizer of $g_z$. For $0<z<1$, i.e.\ in high temperatures, we have $\bar{m}=0$, whereas at $z>1$, $m=0$ is a local maximum of $g_z$ instead, and $\bar{m} \neq 0$ satisfies $\bar{m} = \tanh \left( z \bar{m} \right) $. This is a hallmark of a second order phase transition, with the critical point at $z_{\mathrm{c}}= 1$ indicating a singularity in the free energy density $\mathfrak{f}_N(z)$.

\subsection{Finding the critical exponents}

The singularity in the free energy is smoothed out when taking into account finite size corrections in $N$. To extract the CEs, one typically expands $g_z$ for $z\uparrow z_{\mathrm{c}}$ around $\bar{m} =0 $, leading to a Landau theory with $m$ being the order parameter.  We avoid using here the ideas of the Landau theory altogether. Instead, we infer the CEs from the quadratic (Gaussian) expansion in $m$ only. Note that for the Landau theory, we would need to expand at least to order $m^4$, such that stability of the Landau theory is ensured. 

In the large $N$ limit and for fixed $z<z_{\mathrm{c}}$, the partition function can be evaluated by 
\begin{eqnarray}
\label{eq:ZN}
    {\cal Z}_N (z) &=& \sqrt{\frac{Nz}{2\pi}} e^{N \log 2} \int_{-\infty}^\infty \mathrm{d} m \, e^{- \frac{1}{2}N z(z_{\mathrm{c}}-z) m^2 + O(Nm^4)} 
    \nonumber \\  
    &
    = & 
    e^{N \log 2  - \frac{1}{2}\log \left( z_{\mathrm{c}}-z \right)} (1 + O(1/N))\,,
\end{eqnarray}
taking into account only Gaussian fluctuations around the minimizer $\bar{m} = 0$.
One can hence write the free energy density as
\begin{equation}
\label{eq:fn}
    \mathfrak{f}_N(z) = -\log 2 + \frac{1}{2 N}\log \delta z\, + O\left(1/N^2\right),
\end{equation}
where $\delta z = z_{\mathrm{c}} - z >0$, in this case. Clearly, the free energy in the large $N$ limit displays a logarithmic singularity as $z \uparrow z_{\mathrm{c}} $.

In the case of a continuous phase transition, the free energy density in the vicinity of the critical point is composed of an analytic and a singular contribution:
\begin{eqnarray}
\label{eq:sing}
\mathfrak{f}_N(z) &=&  \mathfrak{f}_0(z) + N^{-\lambda} \mathfrak{f}_{\text{s}}\left(\delta z N^\theta\right ) 
\\ \nonumber
&+& \text{ subleading corrections}.
\end{eqnarray}
Here, the singular scaling function $\mathfrak{f}_{\text{s}}$ can include contributions from the omitted higher-order terms in~\eqref{eq:fn}, which become comparable to the $O(1/N)$ term in the critical window. Nevertheless, even without knowing these terms explicitly, we still can immediately infer from~\eqref{eq:fn} that $\lambda = 1$ here. To find $\theta$, we demand that the second derivative of the singular scaling function in~\eqref{eq:sing} with respect to $z$ becomes comparable to the leading order contribution in the critical window, i.e.\ when we take $\delta z N^\theta = \text{const}$. This is just a restatement of the Ginzburg criterion, implying
\begin{align}
\label{eq:ginz}
     \lambda = 2 \theta.
\end{align}
In the present example, this implies $\lambda=1,\theta =1/2$. To summarize, by evaluating just the Gaussian fluctuations around $\bar{m}=0$ at fixed $z < z_{\text{c}}$ and using the Ginzburg criterion~\eqref{eq:ginz}, we have found the CEs of the CW model. This is the content of the Gaussian fluctuation method. 

In this particular toy model, we note that we can also determine the full singular scaling function, confirming the prediction for the CEs above, see appendix~\ref{sec:appendix-cw}.

Note that we have skipped deriving scaling relations that involve the order parameter, or an external field coupled to the order parameter. These exponents are not included in our theory. We postpone discussion of these CEs and their relations to $\lambda,\theta$ to section~\ref{sec:Summary}.  

In summary, the Gaussian fluctuation method introduced here indicates that one can obtain the CEs by expanding the fluctuations to the Gaussian level, isolating their contribution to the singularity, and then using the scaling relation found from the Ginzburg criterion. In what follows, we now employ these ideas in weak-noise Langevin equations~\eqref{eq:Langevin equation}. 

\section{
Classification of critical exponents
\label{Sec:General theory} } 

We now transition from the framework of equilibrium statistical mechanics to nonequilibrium statistical field theories, represented by~\eqref{eq:Langevin equation} or equivalently~\eqref{eq:path prob h}. While the extraction of CEs for systems undergoing second-order DPTs within statistical field theories presents greater technical challenges compared to the CW model, the core concepts remain applicable. Specifically, the Gaussian fluctuation method, outlined in the previous section by focusing on Gaussian corrections to the saddle-point solution to extract the CEs, still applies.

Building on~\eqref{eq:path prob h} and~\eqref{eq:z-condition-gen}, we previously established that in the small noise limit and close to the critical~$z_{\text{c}}$ of a second-order DPT, we expect the large deviation function to satisfy~\eqref{eq:I singular scaling form}, where
\begin{equation}
\label{eq:LDF I in the weak noise limit}
    I_0(z) = \min_{\phi\in \mathcal{A}_z} S[\phi] ,
\end{equation}
with $\mathcal{A}_z$ the set of fields $\phi(x,t)$ that realize the event~\eqref{eq:z-condition-gen}. That is, the conditional sum amounting to ${\cal P}(z)$ is dominated by an optimal fluctuation $\bar{\phi}$ such that $z = \int \mathrm{d} x \int \mathrm{d} t \, Z [\bar{\phi} ]$. Solving the minimization problem~\eqref{eq:LDF I in the weak noise limit} is generally a non-trivial partial differential equation, which can exhibit strong nonlinearities. However, it has been demonstrated time and again that theoretical and numerical approaches can lead to the determination of the large deviation function.
Following the approach from section~\ref{Sec:Infinite range Ising}, we perform a perturbative expansion of $ S[\bar{\phi} + \delta \phi] $ up to second order in $\delta \phi$, systematically summing the contributions. Here, our primary focus is on the terms that lead to singular behavior, as these are crucial for understanding the critical properties of the DPT. By isolating these singular contributions, we can effectively capture the relevant CEs.
 
First, recall that when a finite $\delta z = z_{\mathrm{c}} - z > 0$ is considered together with the limit $\varepsilon\rightarrow 0$, $I(z)-I_0(z)$ is of order $\varepsilon$. In this limit, 
\begin{align}
\label{eq:delta h2 variation}
&\exp\left\{- \varepsilon^{-1} I(z)\right\} \nonumber\\
&= 
\int_{\bar{\phi}+ \sqrt{\varepsilon}\delta \phi \in {\cal A}_z} \mathcal{D}  [\delta \phi ] \,  \exp \left\{-\varepsilon^{-1} S\left[\bar{\phi}+ \sqrt{\varepsilon}\delta \phi \right] \right\}
 \nonumber
\\ 
& \simeq \exp\bigg\{ -\varepsilon^{-1}I_0(z) \; - \frac{1}{2} \log \varepsilon 
\nonumber \\
& \qquad \qquad + \log \left(\int \mathcal{D} [\delta \phi]  \, e^{- \frac{1}{2} S_2[\delta \phi]} \right) + O(\varepsilon) \bigg\},
\end{align}
where $S_2[\delta \phi] =  \int \mathrm{d} x \int \mathrm{d}t \, \delta \phi \, \left(\delta^2 S_{\bar{\phi}} / \delta \phi^2\right) \delta \phi$ is the second variation of $S$ at $\bar{\phi}$. Notice that since $\bar{\phi}$ is the solution of a minimization problem, the first variation $S_1[\delta \phi]=  \int \mathrm{d} x \int \mathrm{d}t \; \delta \phi \; \delta S_{\bar{\phi}} / \delta \phi $ vanishes at $\bar{\phi}$ \footnote{For a non-local action, the explicit form of the expressions here would be different. Nevertheless, after applying the Fourier analysis in \eqref{eq:Gaussian form}, the results presented in the following become consistent with a non-local action as well.}.  The boundary conditions for the Gaussian fluctuations $\delta \phi$ in~\eqref{eq:delta h2 variation} are such that 
\begin{equation}
    z = \int \mathrm{d} x \int \mathrm{d} t \, Z[\overline{\phi}+\sqrt{\varepsilon}\delta \phi]
\end{equation}
to leading order. We remark that for nonlinear~$Z$, there may be an additional contribution from~$\delta ^2 Z_{\bar{\phi}} / \delta \phi^2$ to the integrand in~\eqref{eq:delta h2 variation}, which we omit here for readability.

One can always express the fluctuations $\delta \phi$ using the Fourier representation 
\begin{equation}
    \delta \phi(x,t) = \sumint_{\; k,\omega} a_{k,\omega} \exp \left\{i k x + i \omega t \right\}.
\end{equation}
 Different boundary conditions for $\phi(x,t)$ and consequently for $\delta \phi$, depending on the specific system at hand, may lead to different conditions on the allowed frequency components $k,\omega$ and corresponding coefficients $a_{k,\omega}$ in this sum, hence the ambiguity in the summation or integration in $\sumint$. Nevertheless, one need not take into account the details of the Fourier transform of the specific problem due to~\eqref{eq:z-condition-gen} for the following general calculation. Instead, we only distinguish between three distinct cases:
\begin{enumerate}
    \item[A.] $k$ is discrete and $\omega$ is continuous (or vice versa), i.e.\ $d_\perp = 1$ continuous frequency.
    \item[B.] $k,\omega$ are both discrete, i.e.\ $d_\perp = 0$ continuous frequencies.
    \item[C.] $k,\omega$ are both continuous, i.e.\ $d_\perp = 2$ continuous frequencies. 
\end{enumerate}
We will argue that the above classification determines the CEs $\lambda,\theta$. 
It is important to stress that the \textit{effective} number $d_\perp$ of continuous frequencies is the deciding factor here. In order to illustrate the general method in the simplest setting possible and avoid treating too many different cases, the reader should have in mind a situation where the  constraint~\eqref{eq:z-condition-gen} depends on the full field configuration in space and time in the following. This means that both frequencies $k$ and $\omega$ are actually present in the problem, and whether they are discrete or continuous depends on the spatial domain and time interval size being finite or infinite.
It is easy to conceive of other scenarios: Consider for instance a Langevin equation~\eqref{eq:Langevin equation} with an equilibrium or nonequilibrium steady state distribution, a constraint~\eqref{eq:z-condition-gen} that depends only on the field at time~$T$, and take the limit $T \to \infty$. Then, even though the problem setup includes time, and in fact a continuous range of frequencies $\omega$, clearly the effective number of continuous frequencies $d_\perp$ can only depend on $k$ being continuous or not, and hence $d_\perp \in \{0, 1 \}$. We will not consider this and other special cases in this section for clarity of the exposition.

Now, for the quadratic expansion in $\delta \phi$ around the optimal fluctuation $\bar{\phi}$, up to a linear transformation in the~$a_{k,\omega}$ terms, we can always rewrite~\eqref{eq:delta h2 variation} using the Fourier decomposition as   
    \begin{equation}
    \label{eq:Gaussian form}
        \delta I(z)
        \simeq - \varepsilon \log \int \bigg( \prod_{k,\omega} \mathrm{d} a_{k,\omega } \mathrm{d} a_{k,\omega }^* \bigg)  e^{- \sumint_{\; k,\omega} |a_{k,\omega}|^2 g_z (k,\omega)}\,,
    \end{equation}
    where $\delta I(z)= I(z) - I_0(z)$. Importantly, the (possibly linearly transformed) Fourier decomposition  diagonalizes the second variation, so that different modes can be treated independently. Here, $g_z$ is a non-negative function for~$\delta z>0$, as we have assumed that~$\bar{\phi}$ is indeed a true and nondegenerate local minimum. We note that actually finding a diagonal decomposition of the second variation in practice can be difficult. While it is indeed possible in both examples we consider in sections~\ref{sec:KPZ} and~\ref{sec:MFT}, there are other cases where it may not be straightforward. We further comment on this point in section~\ref{sec:Summary}.

Our goal is to rearrange the right hand side of~\eqref{eq:Gaussian form} into the anomalous and subleading correction terms in~\eqref{eq:I singular scaling form}. In order to do that, we stress that the Langevin equation~\eqref{eq:Langevin equation} is typically a hydrodynamic equation. Hence, there are microscopic cutoffs in time and distance. This translates to having cutoffs for the Fourier frequencies $k_{\text{max}},\omega_{\text{max}}$. While it is impossible to infer the value of the cutoffs from the hydrodynamic theory, universality of the second order DPT states that the 
singular scaling function~$I_{\text{s}}$ cannot depend on these cutoffs. This is because the DPT involves a macroscopic change in the properties of the system due to fluctuations, which cannot be attributed to microscopic details. In simpler terms and for the CW toy example: the inter-spin coupling, i.e.\ the specific value of $J$ affected by the neighboring spins distance, cannot affect the CE values.

All the above implies that~\eqref{eq:Gaussian form} can be brought to the form 
\begin{equation}
\label{eq:log sum}
       \delta I(z) \simeq \varepsilon \sumint_{\; k,\omega} \log g_z(k,\omega) + \varepsilon \mathcal{C}(k_{\text{max}},\omega_{\text{max}})\,,
\end{equation}
    where we absorb all non-singular, subleading and cutoff-dependent terms into the expression $\mathcal{C}(k_{\text{max}},\omega_{\text{max}})$. 

At the transition $z \uparrow z_{\text{c}}$, we expect there to be (at least, and often exactly) one frequency tuple $\left( k_0,\omega_0\right)$ such that $g_{z_{\mathrm{c}}}(k_0,\omega_0) =0 $, as a precursor of the transition of $\bar{\phi}$ from a local minimum to a saddle point. This is Le Chatelier's principle \cite{shpielberg2016chatelier}. The coefficient function $g_z$ itself can reasonably be assumed to be an analytic function for $\delta z>0$. Hence, we consider
\begin{equation}
\label{eq: k0 omega0 f}
g_{z\rightarrow z_{\mathrm{c}}} (k_0,\omega_0) \simeq  A \left( \delta z \right)^n    
\end{equation}
 to leading order in $\delta z$, where $A>0$, and
 $n$ is a positive integer, typically $n = 1$.
 At this point, we will address the three different possibilities for the Fourier frequencies $k$ and $\omega$.

\subsection{One continuous Fourier frequency component
\label{subsec:one cont Fourier}}

Consider the case of a discrete frequency component $k$ and continuous frequency component $\omega$ (without loss of generality). We are interested in evaluating the sum in~\eqref{eq:log sum}, and single out the cutoff-independent singular terms appearing in the limit of $\delta z \downarrow 0$. Taking the continuum limit for $\omega$, such that the frequency gap $\Delta \omega \to 0$, we are interested in evaluating $\frac{1}{\Delta \omega} \sum_k \int_{-\omega_{\text{max}}}^{\omega_{\text{max}}} \mathrm{d} \omega \log g_z(k,\omega)   $. Here $\Delta \omega \rightarrow 0$ depends on the details of the problem at hand, but should not scale with  $\varepsilon$.

In~\eqref{eq: k0 omega0 f}, we have assumed that $g_{z\uparrow z_c}(k, \omega) \rightarrow 0$ only for $\left(k,\omega\right) = \left(k_0,\omega_0\right)$. Since the frequencies $k$ are discrete, we can surmise that there exists a constant $\mathcal{D}$, such that  $g_z(k\neq k_0,\omega) > \mathcal{D} > 0$ for any $\omega$ when $\delta z>0$, and for any $k \neq k_0$. Therefore, the summation over all frequencies $k\neq k_0$ is a non-singular term (that may depend on the cutoff) and can be absorbed into $\mathcal{C}$. 

\begin{figure}
    \centering
    \includegraphics[width=0.85\linewidth]{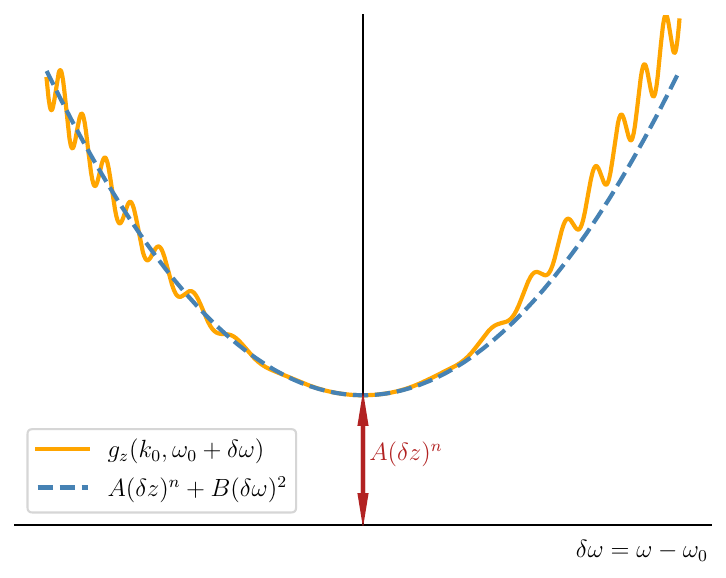}
    \caption{ $g_z(k_0,\omega)$ is typically a non-trivial function. Assuming $k$ are discrete frequencies and $\omega$ continuous frequencies, as $z\uparrow z_{\text{c}}$ and $\omega \approx \omega_0$, the function can be approximated as seen in the figure. Contributions to the singular scaling function can only come from regions where $g_z(k, \omega) \rightarrow 0$. 
    }
    \label{fig:gz-approx}
\end{figure}

Indeed, the contribution to the singular terms arises from the $k_0$ frequency. Moreover, the significant contribution of the $k_0$ frequency comes from the vicinity of~$\omega$ around~$\omega_0$, see Fig.~\ref{fig:gz-approx}. This suggests considering a  Taylor expansion of $g_z(k_0, \omega)$ around $\omega=\omega_0$, which is $g_z(k_0,\omega) = A \left( \delta z \right)^n + B \left( \delta \omega \right)^2 + C \left( \delta \omega \right)^3 + \dots $. Here, the positivity of $g_z$ requires a vanishing linear term in $\delta \omega = \omega-\omega_0$, and $B>0$. 
Note that additional mixed terms $\propto (\delta z)^n \, \delta \omega$ and higher orders may appear in the Taylor expansion of $g_z(k_0, \omega)$ for certain field theories, such as the KPZ example from section~\ref{sec:KPZ}, but these will not change the following argument, and hence we omit them here.

Now, we define $y=\sqrt{B/A} \, \delta \omega / \left( \delta z \right)^{n/2}  $ and calculate the contribution of the $k_0$ frequency 
\begin{align}
\label{eq:z integral}
 &   \intop^{\omega_{\text{max}} } _{-\omega_{\text{max}}} 
    \mathrm{d} \omega \, \log g_z(k_0,\omega) =
   \sqrt{\frac{A \left(\delta z\right)^n}{B}} \int_{y_{\text{min}}}^{y_{\text{max}}} \mathrm{d} y  \log \left(  A \left( \delta z \right)^n \right) \nonumber
    \\ 
     & \quad \quad + \sqrt{\frac{A \left(\delta z\right)^n}{B}} \int_{y_{\text{min}}}^{y_{\text{max}}} \mathrm{d} y \log \left( 1+ y^2 +O\left( (\delta z)^{n/2} y^3\right)\right).
\end{align}
The first term is indeed singular, but still depends on the cutoff limits. Therefore, it cannot contribute to the singular scaling function as it would violate universality. This singular contribution turns out to be subleading as can be expected.   
We  notice that the limits of the $y$ integral approach $(y_{\text{min}}, y_{\text{max}}) \to (-\infty,\infty) $  as $\delta z \downarrow 0$. This implies that we can safely take the second term in the right hand side of~\eqref{eq:z integral} to be 
\begin{equation}
    \int^{y_{\text{max}}} _{y_{\text{min}}} dy \, \log \left(1+y^2 \right) =  2\pi + \mathcal{B}(\omega_{\text{max}}).
\end{equation}
The $\mathcal{B}(\omega_{\text{max}})$ term above depends on the cutoffs and will be absorbed into $\mathcal{C}$. Hence, summing up all contributions, a single singular contribution exists, leading to  
\begin{equation}
\label{eq:delta I exact single continouos}
    \delta I(z) = \varepsilon \frac{2\pi}{\Delta \omega} \sqrt{\frac{A}{B}} \left( \delta z \right)^{n/2} + \varepsilon \mathcal{C}\left(k_{\text{max}},\omega_{\text{max}}\right) .
\end{equation}
All in all, using~\eqref{eq:delta I exact single continouos} and~\eqref{eq:I singular scaling form} we find 
\begin{equation}
\label{eq:scaling n}
    \varepsilon^\lambda I_{\text{s}} \left(\delta z / \varepsilon^\theta \right)  \propto \varepsilon \left(\delta z \right)^{n/2} . 
\end{equation}
This is sufficient to find a scaling relation. Setting $\delta z = \varepsilon^\theta$, we find 
\begin{equation}
    \lambda = 1 + \theta n /2. 
\end{equation}
Given that typically $n=1$, we need one more equation to determine both the CEs $\lambda$ and $\theta$. 

To obtain the second equation, we use the second derivative test or the Ginzburg criterion as discussed above. Here, we demand that the second derivative of the contribution from $I_{\text{s}}$ with respect to $z$, will be comparable to the leading order contribution. This simply implies $ \lambda= 2\theta$. This description corresponds to a second order DPT. If we have a continuous DPT of order~$G_{\mathrm{c}}$ instead, with $G_{\mathrm{c}}$ an integer
larger than one, then we deduce $\lambda = G_{\mathrm{c}} \theta $. The two scaling rules lead to 
\begin{equation}
    \lambda = \frac{2G_{\mathrm{c}}}{2G_{\mathrm{c}}-n} , \quad \theta =  \frac{2}{2G_{\mathrm{c}}-n}\,.
    \label{eq:crit-expo-one-continous}
\end{equation}
In particular, for the simplest case where $n=1,G_{\mathrm{c}}=2$, we find $\lambda= 4/3$, $\theta = 2/3$. In section~\ref{sec:KPZ} and section~\ref{sec:MFT} we will show several examples corresponding to this case; in particular, this is the relevant case for the large deviations of the current in a set of boundary driven and periodic MFT processes, as well as the spatially averaged height of a KPZ interface on an infinitely large ring.

We remark that the order of the transition $G_{\mathrm{c}}$ depends on the structure of the coefficient function $g_z$ and is hence not independent of the exponent $n$ in the Taylor expansion~\eqref{eq: k0 omega0 f}. In particular, within the scope of Landau theory, $G_{\mathrm{c}}$ depends on higher-order terms in the Landau free energy. Hence, since we only consider Gaussian fluctuations around the optimal fluctuation here, we will assume that~$G_{\mathrm{c}}$ is given and determined through other means (e.g.\ through numerical minimization of the action~\eqref{eq:action-func} at $z > z_{\text{c}}$).

\subsection{No continuous Fourier frequency components}
\label{sec:no-cont-components}

Here we consider the case where the $k,\omega$ frequencies are both discrete.  
Starting from~\eqref{eq:Gaussian form}, only the contribution from the $\left(k_0,\omega_0 \right)$ frequencies lead a singular contribution at the $\delta z\downarrow 0$ limit. Integrating over the right hand side of~\eqref{eq:Gaussian form}, we find 
\begin{equation}
\label{eq:two discrete scaling}
 \delta I(z)    = \varepsilon \log g_z (k_0,\omega_0) + \varepsilon \mathcal{C}(k_{\text{max}},\omega_{\text{max}}). 
\end{equation}
For $z\uparrow z_{\mathrm{c}}$, we again expect $g_{z\uparrow z_{\mathrm{c}}}(k_0,\omega_0) = A (\delta z)^n$.  Taking~\eqref{eq:two discrete scaling} and the scaling of $g_{z\uparrow z_{\mathrm{c}}}(k_0,\omega_0)$, this leads to 
\begin{equation}
\label{eq:discrete scaling}
    \varepsilon^{\lambda} I_{\text{s}}\left(\delta z / \varepsilon^\theta \right) \simeq  \varepsilon \log \left( \delta z/\varepsilon^\theta \right), 
 \end{equation}
where we have recovered a term $\varepsilon \log \varepsilon^\theta$ from $\mathcal{C}$, and a factor of~$n$ is omitted as it does not affect the CEs. 
Setting  $\delta z = \varepsilon^\theta$ in~\eqref{eq:discrete scaling},  we find $\lambda = 1$. The Ginzburg criterion 
$\lambda =G_{\mathrm{c}} \theta  $  implies 
\begin{equation}
\theta=1/G_{\mathrm{c}}, \quad  \lambda =1. 
\end{equation}
This is the case already demonstrated in the CW model in section~\ref{Sec:Infinite range Ising}, where $G_{\mathrm{c}}=2$, such that $\theta = 1/2$, $\lambda = 1$. We will also demonstrate this scaling in \ref{sec:KPZ} for short-time large deviations of the KPZ equation on a ring of finite length.

\subsection{Two continuous Fourier frequencies}
\label{sec:two-cont-components}

Finally, we turn to the case where the two Fourier frequencies $k,\omega$ are both continuous. As might be expected by this point, $g_z$ vanishes in the limit of $\delta z\downarrow 0$ only in the vicinity of $\left(k_0,\omega_0 \right)$. Following section~\ref{subsec:one cont Fourier}, we integrate over the frequencies $k,\omega$. Moreover, the relevant singular terms can only come from the integration in the vicinity of the frequencies $(k_0,\omega_0)$. Therefore, we expand $g_z$ around  $(k_0,\omega_0)$, and integrate this contribution. Namely,       
\begin{align}
\label{eq:two cont. integral}
    \delta I(z) = \varepsilon {\cal C}  + \frac{\varepsilon}{\Delta \omega \Delta k} \int_{-k_{\text{max}}}^{k_{\text{max}}} \mathrm{d} k \int_{-\omega_{\text{max}}}^{\omega_{\text{max}}} \mathrm{d} \omega \, \log g_{z\uparrow z_{\mathrm{c}}}(k, \omega)
\end{align}
where
\begin{align}
    g_{z\uparrow z_{\mathrm{c}}}(k, \omega) =  A (\delta z)^n + \frac{B}{2} \left(\delta \omega \right)^2 +  \frac{C}{2} \left( \delta k \right)^2 + D \left(\delta k\right) \left( \delta \omega \right),
\end{align}
where $B,C>0$ with $CB-D^2> 0$ to ensure the positivity of $g_z$. Also, we have defined $\delta \omega = \omega - \omega_0$ and $\delta k = k  - k_0$. The integration is relevant close to  $(k_0,\omega_0)$, as the rest gives a non-singular contribution to be absorbed into $\mathcal{C}$.

The integral in~\eqref{eq:two cont. integral} is slightly more cumbersome, but similar to~\eqref{eq:z integral}. The leading order singular term gives
\begin{equation}
    \varepsilon^\lambda I_{\text{s}} \left(\delta z / \varepsilon^\theta  \right)  \propto \varepsilon  \left(\delta z\right)^n \log \left(\delta z / \varepsilon^\theta \right) . 
\end{equation}
The Ginzburg criterion gives $\lambda = G_{\mathrm{c}} \theta $ again, and from the scaling relation of $I_{\text{s}}$, we find $\lambda = 1+\theta n$. In total, this results in 
\begin{equation}
   \lambda  =  \frac{G_{\mathrm{c}}}{G_{\mathrm{c}}-n} ,  \quad \theta =  \frac{1}{G_{\mathrm{c}}-n}.
   \label{eq:crit-expo-two-continous}
\end{equation}
The simplest case $n = 1$ and $G_{\text{c}} = 2$ gives $\lambda = 2$ and $\theta = 1$.
In this paper, we do not present a specific example where both two Fourier frequencies are continuous. To find an example, one may consider either higher dimensions or conditioning on more than one fluctuating field $\phi$.  

In what follows, we demonstrate our findings in multiple examples, by recalling known continuous DPTs in the literature and extracting the corresponding CEs, which will be seen to match the general predictions.

\section{Short-time large deviations of the Kardar--Parisi--Zhang equation\label{sec:KPZ}}

The KPZ equation, proposed in~\cite{kardar-parisi-zhang:1986} as a continuum model of nonlinear interface growth, plays a prominent role in nonequilibrium statistical physics because of its appearance as the limiting equation of a large number of microscopic models within its universality class~\cite{corwin:2012,takeuchi2018appetizer}. Among the many statistical properties of the KPZ equation in $1+1$ dimensions that have hence been studied in the literature~\cite{quastel2015one,prolhac2024kpz}, we focus on short-time large deviations here, i.e.\ characterizing the probability and shape of rare growth events of the interface after a short time~$T$, when starting from a given initial condition. In a suitable choice of units, where $T = 1$ is fixed instead, this corresponds to the small-noise Langevin setting of equation~\eqref{eq:Langevin equation} for the interface height $h = h(x,t)$ (see e.g.~\cite{meerson-katzav-vilenkin:2016}). Depending on the initial condition, spatial boundary conditions, and observable~$Z$ in~\eqref{eq:z-condition-gen}, various DPTs have been found and analyzed in this system, such as: a continuous DPT associated with mirror symmetry breaking for a KPZ interface on the real line with stationary initial condition and one-point height observable $\int \mathrm{d} x \int \mathrm{d} t \, Z[h] = h(x = 0, t = 1)$~\cite{janas-kamenev-meerson:2016,krajenbrink-le-doussal:2017,krajenbrink-doussal-prolhac:2018,smith-kamenev-meerson:2018,hartmann-meerson-sasorov:2021,krajenbrink-le-doussal:2022}, a first-order DPT for a KPZ interface on the half-line with reflecting boundary and the one-point height observable at a shifted point~\cite{asida-livne-meerson}, and DPTs of different orders for the one-point and spatially averaged height of a KPZ interface with flat initial condition when defined on a ring~\cite{smith-meerson-sasorov:2018,schorlepp-grafke-grauer:2023,schorlepp-sasorov-meerson:2023}. Thus, the short-time limit of the KPZ equation constitutes an ideal testing ground for the general theory described in the previous section. We will consider the short-time large deviations of the spatially averaged height of a KPZ interface on a ring here, whose phase diagram has recently been studied in~\cite{schorlepp-grafke-grauer:2023,schorlepp-sasorov-meerson:2023}.

\subsection{Large deviations of the spatially averaged height on ring of finite length $\ell$}

Consider a one-dimensional interface $h(x,t)$ on $[0,\ell] \times [0,1]$ evolving according to
\begin{align}
    \partial_t h = \partial_{xx} h + \frac{1}{2} \left(\partial_x h \right)^2 + \sqrt{\varepsilon} \xi(x,t)
    \label{eq:kpz}
\end{align}
in non-dimensionalized units. The interface starts from a flat initial condition $h(x, 0) = 0$ and is defined on a ring of length $\ell$, i.e.\ with periodic boundary conditions in space. Following~\cite{schorlepp-grafke-grauer:2023,schorlepp-sasorov-meerson:2023}, we analyze the rare event statistics of the spatially averaged surface height
\begin{align}
    \bar{H} =\frac{1}{\ell}\int_0^\ell \mathrm{d} x \; h(x,1)
    \label{eq:kpz-average-height}
\end{align} 
at final time $t = 1$. 

The optimal fluctuation $\bar{h}$ is the solution of~\cite{fogedby:1999,kolokolov-korshunov:2009,meerson-katzav-vilenkin:2016}
\begin{align}
\begin{cases}
    \partial_t \bar{h} = \partial_{xx} \bar{h} + \frac{1}{2} \left(\partial_x \bar{h} \right)^2 + \bar{p}\,,\\
    \partial_t \bar{p} = -\partial_{xx} \bar{p} + \partial_x \left(\bar{p} \partial_x \bar{h} \right)\,,
\end{cases}
\label{eq:opt-fluc-kpz}
\end{align}
where $\bar{p}$ is the conjugate momentum density. The boundary conditions are $\bar{h}(x,0) = 0$, $\int_0^\ell \mathrm{d}x \; \bar{h}(x,1) = \ell \bar{H}$ and $\bar{p}(x,1) = \Lambda = \text{const}$ for a Lagrange multiplier $\Lambda = \Lambda(\bar{H})$, chosen in such a way as to ensure the final-time constraint $\int_0^\ell \mathrm{d}x \; \bar{h}(x,1) = \ell \bar{H}$ holds. An obvious solution to~\eqref{eq:opt-fluc-kpz} with these boundary conditions is given by the uniformly growing profile $\bar{h}(x,t) = \bar{H} t$ with momentum $\bar{p}(x,t) = \Lambda = \bar{H}$ and quadratic action $I_0(\bar{H}) = S [\bar{h} ] = \ell \bar{H}^2 / 2$. As shown in~\cite{schorlepp-sasorov-meerson:2023} through linear stability analysis, the uniformly growing profile $\bar{h}$ becomes unstable as $\bar{H} \uparrow \bar{H}_{\text{c}}$, with the critical average height $\bar{H}_{\text{c}} = \bar{H}_{\text{c}}(\ell) > 0$ as the smallest nontrivial real solution of
\begin{align}
    \tan \left(k \sqrt{\bar{H}_{\text{c}} - k^2} \right) + k^{-1} \sqrt{\bar{H}_{\text{c}} - k^2} = 0
\label{eq:stability-kpz}
\end{align}
among the allowed frequencies $k = 2 \pi m / \ell$ with $m = 1, 2, 3, \dots$ This signals the appearance of a continuous DPT of second order $G_{\text{c}} = 2$~\cite{schorlepp-grafke-grauer:2023,schorlepp-sasorov-meerson:2023} at $\bar{H}_{\text{c}}$, the critical exponents of which we will calculate explicitly in the following. Note that, as analyzed in detail in~\cite{schorlepp-sasorov-meerson:2023}, at large $\ell \to \infty$ the second-order DPT at $\bar{H}_{\text{c}}$ is in fact ``masked'' by another (weakly) first order DPT from the uniform profile to a solitonic solution of~\eqref{eq:opt-fluc-kpz} which occurs at a smaller $\bar{H}$. Nevertheless, we can still compute the critical behavior of the system in the vicinity of the uniform profile as $\bar{H} \uparrow \bar{H}_{\text{c}}$, even at large $\ell$. We do not address any of the potential critical points in the $\ell$-$\bar{H}$ phase diagram that arise due to the first order transition line here.\\

Following the general strategy from section~\ref{Sec:General theory}, we expand the action~\eqref{eq:action-func} around the profile~$\bar{h}$ up to order~$\varepsilon$ with $h(x,t) = \bar{h}(x,t) + \sqrt{\varepsilon} \delta h(x,t)$. For finite domain size~$\ell$, we expect to recover case B treated in section~\ref{sec:no-cont-components}, and as $\ell \to \infty$, we should get case A from section~\ref{subsec:one cont Fourier}. We remark that in principle, the contribution of Gaussian fluctuations around $\bar{h}$ was already evaluated in~\cite{schorlepp-grafke-grauer:2023} using Riccati equations~\cite{schorlepp-grafke-grauer:2021,bouchet-reygner:2022,grafke-schaefer-vanden-eijnden:2024} for the stochastic heat equation (SHE), which is related to the KPZ equation~\eqref{eq:kpz} through a Cole--Hopf transformation. Our goal here is to show how the method from section~\ref{Sec:General theory} works in a concrete example. We will compare with the available result from~\cite{schorlepp-grafke-grauer:2023} in the end. In the present example, the Gaussian fluctuations $\delta h(x,t)$ need to satisfy $\delta h(x,0) = 0$ and $\int_0^\ell \mathrm{d} x \; \delta h(x,1) = 0$, in addition to periodic boundary conditions in space. Without loss of generality, we decompose the fluctuations into $\delta h(x,t) = \delta h_0(x,t) + t \delta h_1(x)$, now with $\delta h_0(x,0) = \delta h_0(x,1) = 0$ and $\int_0^\ell \mathrm{d} x\; \delta h_1(x) = 0$, in order simplify the analysis in Fourier space. Expanding the action~\eqref{eq:action-func}, we obtain
\begin{align}
    &\tfrac{1}{\varepsilon} S\left[\bar{h} + \sqrt{\varepsilon} \delta h \right] - \tfrac{1}{\varepsilon} I_0\left(\bar{H} \right) \nonumber\\
&= \tfrac{1}{2} \int_0^1 \mathrm{d}t \int_0^\ell \mathrm{d}x \; \bigg(\delta h_0 \left[-\partial_{tt} + \bar{H} \partial_{xx} + \partial_{xxxx} \right] \delta h_0 \nonumber\\
&\quad + \delta h_1 \left[1 + \left(\bar{H}t^2 - 2 t \right) \partial_{xx} + t^2 \partial_{xxxx} \right] \delta h_1 \nonumber\\
&\quad + 2 \delta h_1 \left[\partial_t - t \partial_t \partial_{xx} - \partial_{xx} + t \partial_{xxxx} + \bar{H} t \partial_{xx} \right] \delta h_0\bigg)
\label{eq:second-var-kpz-real-space}
\end{align}
up to terms of order $O\left(\varepsilon^{1/2}\right)$. Here, we already omitted the vanishing term $\int_0^1 \mathrm{d}t \int_0^\ell \mathrm{d}x \; \delta h_0 \partial_t \partial_{xx} \delta h_0 = - \tfrac12\int_0^1 \mathrm{d}t \int_0^\ell \mathrm{d}x \; \partial_t \left(\partial_x \delta h_0 \right)^2 = 0$ in the expansion. We let
\begin{align}
    \delta h_0(x,t) &= \sum_{\substack{k \in 2 \pi \mathbb{Z} / \ell\\ \omega \in \pi \mathbb{N}}} a_{k, \omega} \exp \left\{i k x \right\} \sin\left(\omega t \right)\\
    \delta h_1(x) &= \sum_{k \in 2 \pi \mathbb{Z} / \ell} b_k \exp \left\{i k x \right\} 
\end{align}
with $a_{-k, \omega} = a_{k, \omega}^*$, $b_{-k} = b_k^*$, and $b_0 = 0$, and $\mathbb{N} = \{1,2,3,\dots\}$ to account for the boundary conditions of the fluctuations. Note that for finite~$\ell$, we have both frequencies $k$ and $\omega$ discrete here. The second variation of the action~\eqref{eq:second-var-kpz-real-space} in terms of these Fourier coefficients becomes
\begin{align}
&\tfrac{1}{2} S_2[\delta h] = \frac{\ell}{2} \bigg( \sum_{\substack{k \in 2 \pi \mathbb{Z} / \ell\\ \omega \in \pi \mathbb{N}}} \big( \left| a_{k, \omega} \right|^2 g_{\bar{H}}^{(a)}(k,\omega) \nonumber\\
&+ \left(a_{k,\omega} b_k^* + a_{k, \omega}^* b_k \right) g_{\bar{H}}^{(ab)}(k,\omega) \big)
    + \sum_{k \in 2 \pi \mathbb{Z} / \ell} \left| b_k \right|^2 g_{\bar{H}}^{(b)}(k) \bigg)
    \label{eq:second-var-kpz-fourier}
\end{align}
with real-valued coefficients
\begin{align}
\begin{cases}
    g_{\bar{H}}^{(a)}(k,\omega) &=  \tfrac{1}{2} \left[\omega^2 - k^2 \left[ \bar{H} - k^2 \right] \right]\\
    g_{\bar{H}}^{(ab)}(k,\omega) &= \tfrac{k^2 \cos \omega}{\omega} \left[ \bar{H} - k^2\right]\\
    g_{\bar{H}}^{(b)}(k) &= 1 + \left(1 - \tfrac{1}{3} \bar{H} \right) k^2 + \tfrac{1}{3} k^4
\end{cases}\,.
\end{align}
Evaluating the Gaussian functional integral~\eqref{eq:delta h2 variation} given by
\begin{align}
        I\left(\bar{H}\right) - I_0\left(\bar{H}\right) = \delta I\left(\bar{H}\right)
        \simeq -\varepsilon \log  \int {\cal D}\left[\delta h \right]  \, e^{- \tfrac{1}{2} S_2[\delta h]}
\end{align}
in the Fourier representation~\eqref{eq:second-var-kpz-fourier} leads to
\begin{align}
   \delta I\left(\bar{H}\right) &\simeq \varepsilon \bigg[\sum_{\omega \in \pi \mathbb{N}} \log \left( \tfrac{1}{2} g_{\bar{H}}^{(a)}(0, \omega) \right) + \sum_{\substack{k \in 2 \pi \mathbb{N} / \ell\\ \omega \in \pi \mathbb{N}}}  \log  g_{\bar{H}}^{(a)}(k, \omega)  \nonumber\\
    &\; + \sum_{k \in 2 \pi \mathbb{N} / \ell} \log \bigg(g^{(b)}_{\bar{H}}(k)- \sum_{\omega \in \pi \mathbb{N}} \tfrac{\left( g_{\bar{H}}^{(ab)}(k,\omega) \right)^2}{g_{\bar{H}}^{(a)}(k, \omega)}\bigg) \bigg]\,.
    \label{eq:delta-i-kpz-result-coeff}
\end{align}
In particular, the inner sum in the last term of~\eqref{eq:delta-i-kpz-result-coeff} evaluates to
\begin{align}
&\sum_{\omega \in \pi \mathbb{N}} \tfrac{\left( g_{\bar{H}}^{(ab)}(k,\omega) \right)^2}{g_{\bar{H}}^{(a)}(k, \omega)} = \sum_{m = 1}^\infty \frac{2 k^4 \left(\bar{H} - k^2 \right)^2}{\pi^2 m^2 \left[\pi^2 m^2 - k^2 \left(\bar{H} - k^2 \right) \right]} \nonumber\\
&= 1 - \tfrac{1}{3} k^2 \left(\bar{H} - k^2 \right) - k \sqrt{\bar{H} - k^2} \cot \left( k \sqrt{\bar{H} - k^2} \right)\,,
\end{align}
and similarly the contribution of each $k$ in the second term in~\eqref{eq:delta-i-kpz-result-coeff} can be written as
\begin{align}
    &\sum_{\omega \in \pi \mathbb{N}}  \log g_{\bar{H}}^{(a)}(k, \omega) = \sum_{\omega \in \pi \mathbb{N}}  \log \left(\tfrac{\omega^2}{2} \right) + \sum_{m = 1}^\infty \log \left(1 - \tfrac{k^2 \left(\bar{H} - k^2 \right)}{m^2 \pi^2} \right) \nonumber\\
    &= \sum_{\omega \in \pi \mathbb{N}}  \log \left(\tfrac{\omega^2}{2} \right) + \log \left(\tfrac{\sin \left(k \sqrt{\bar{H} - k^2} \right)}{k \sqrt{\bar{H} - k^2} } \right)\,.
\end{align}
All in all, we obtain
\begin{align}
    &\delta I(\bar{H}) \simeq \varepsilon \bigg[\overbrace{\sum_{\omega \in \pi \mathbb{N}} \log\left( \tfrac{\omega^2}{4} \right) + \sum_{\substack{k \in 2 \pi \mathbb{N} / \ell\\ \omega \in \pi \mathbb{N}}}  \log \left(\tfrac{\omega^2}{2} \right)}^{={\cal C}\left(k_{\text{max}},  \omega_{\text{max}}\right)} \nonumber \\
    &+ \sum_{k \in 2 \pi \mathbb{N} / \ell} \log \bigg( \underbrace{\tfrac{k^2 \sin \left( k \sqrt{\bar{H} - k^2}\right) +k \sqrt{\bar{H} - k^2} \cos \left( k \sqrt{\bar{H} - k^2}\right)}{k \sqrt{\bar{H} - k^2}}}_{=g_{\bar{H}}(k)} \bigg)\bigg]
    \label{eq:kpz-delta-i-result}
\end{align}
for the leading correction in $\varepsilon$ to $I_0$ for the uniformly growing profile $\bar{h}$ in the KPZ equation~\eqref{eq:kpz} with average height observable~\eqref{eq:kpz-average-height}. It is instructive to compare the result~\eqref{eq:kpz-delta-i-result} to the one obtained in~\cite[Eq.~(C23)]{schorlepp-grafke-grauer:2023} through a different method, namely solving differential Riccati equations~\cite{schorlepp-grafke-grauer:2021,bouchet-reygner:2022,grafke-schaefer-vanden-eijnden:2024} for the SHE. Besides the different choice of units in~\cite{schorlepp-grafke-grauer:2023}, both results agree when accounting for (i) the cut-off dependent ${\cal C}\left(k_{\text{max}},  \omega_{\text{max}}\right)$ here, since the KPZ equation in $1+1$ dimensions, in contrast to the SHE, requires renormalization, and (ii) consequently different $O(1)$ constants that depend on the choice of renormalization and are irrelevant for the critical behavior in the vicinity of $\bar{H}_{\text{c}}$. Regarding the second point, see e.g.~\cite{appert-roland-derrida-lecomte-etal:2008} for an MFT calculation where, similarly, the cut-off dependent terms are determined through comparison with the exact microscopic result.

\begin{figure}
    \centering
    \includegraphics[width=.9\linewidth]{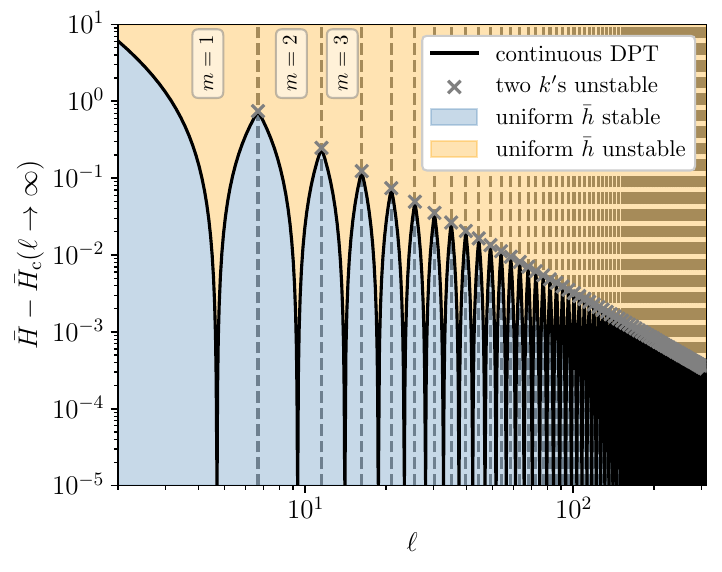}
    \caption{Stability region, according to~\eqref{eq:stability-kpz}, of the uniformly growing interface $\bar{h}(x,t) = \bar{H} t$ for short-time large deviations of the KPZ equation~\eqref{eq:kpz} on a ring of length $\ell$. For better visualization, the vertical axis has been shifted by $\bar{H}_{\text{c}}(\ell \to \infty) \approx 4.60334$. The dashed vertical lines delimit the $\ell$-regions where the frequencies $k = 2 \pi m / \ell$ for $m =1, 2, 3, \dots$, respectively, are the first to become unstable as $\bar{H}$ increases. Hence, at the crosses, $\bar{h}$ becomes unstable to perturbations at two frequencies, corresponding to $m$ and $m+1$, simultaneously.}
    \label{fig:kpz-stability}
\end{figure}

\begin{figure}
    \centering
    \includegraphics[width=0.8\linewidth]{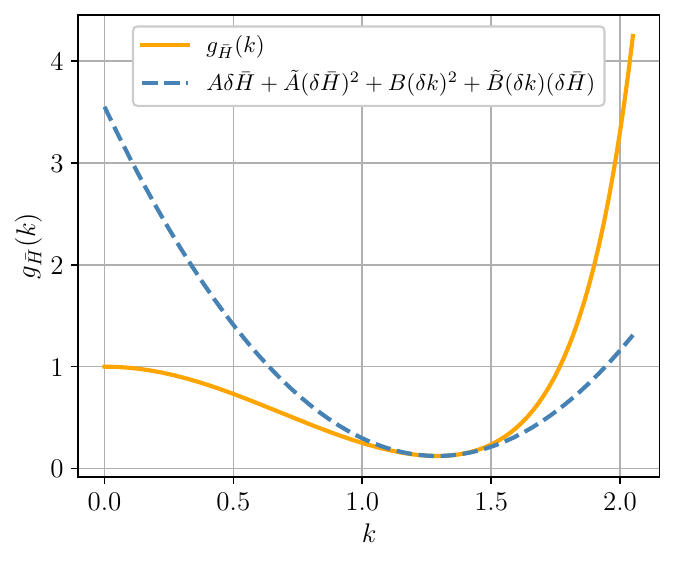}
    \caption{Second-order Taylor expansion at large $\ell$ of the coefficient function $g_{\bar{H}}(k)$ for the KPZ equation~\eqref{eq:kpz}, as defined through~\eqref{eq:kpz-delta-i-result}, around $k_0 \simeq 1.343$ for $\delta \bar{H} = \bar{H}_{\text{c}}(\ell \to \infty) - \bar{H} = 0.1$. The Taylor coefficients are $A \approx 0.626$, $\tilde{A} \approx 0.112$, $B \approx 2.057$, $\tilde{B} \approx 1.068$.}
    \label{fig:g-approx-kpz}
\end{figure}

The result~\eqref{eq:kpz-delta-i-result} also confirms the linear stability criterion~\eqref{eq:stability-kpz}. This criterion gives rise to a complicated structure of the stability region of the uniformly growing interface $\bar{h}(x,t) = \bar{H} t$, which we show in Fig.~\ref{fig:kpz-stability}. The mode~$m$ to become unstable first as~$\bar{H}$ is increased grows with the domain size~$\ell$, with $m \simeq 1.343 \ell / 2 \pi$ and $\bar{H}_{\text{c}} \to 4.603$ as $\ell \to \infty$. At any finite $\ell$, there are either one or two frequencies~$k_0$ which become unstable at the critical $\bar{H}_{\text{c}}(\ell)$. With regard to the Taylor expansion of~$g_{\bar{H}}(k)$ in~\eqref{eq: k0 omega0 f} that determines the CEs, we note that the first derivative of~$g_{\bar{H}}(k)$ in~\eqref{eq:kpz-delta-i-result} at the critical average height and unstable frequencies is given by
\begin{align}
    A = -\left. \tfrac{\partial g_{\bar{H}} (k_0)}{\partial \bar{H}} \right|_{\bar{H} = \bar{H}_{\text{c}}} = -\tfrac{\bar{H}_{\text{c}} + 1}{2\left(\bar{H}_{\text{c}} - k_0^2 \right)} \cos \left( k_0 \sqrt{\bar{H}_{\text{c}} - k^2_0} \right) > 0\,,
\end{align}
indicating that $n = 1$ in the present example. Proceeding as in~\eqref{eq:discrete scaling}, we conclude that $\lambda = 1$ and $\theta = 1/2$ for the CEs at finite $\ell$.

\subsection{The limit $\ell \to \infty$}

As the size $\ell$ of the ring tends to infinity and the allowed frequencies $k$ approach the continuum, it is necessary to take into account the vicinity of the frequency $k_0 \simeq 1.343$ to become unstable first, as explained in the general derivation of section~\ref{subsec:one cont Fourier}. Importantly, the limit $\ell \to \infty$ is not a mere mathematical construction: the domain size $\ell$ in the non-dimensional units used here is given by $\ell \propto L / \sqrt{T}$ in terms of the physical domain size~$L$ and time~$T$~\cite{schorlepp-grafke-grauer:2023,schorlepp-sasorov-meerson:2023}. Hence, when taking the short time limit $T \to 0$ at fixed~$L$, the large $\ell$ limit in non-dimensional units is in fact natural to consider. In the large $\ell$ limit, equation~\eqref{eq:kpz-delta-i-result} becomes
\begin{align}
     \delta I\left(\bar{H}\right) &\simeq \varepsilon \bigg[{\cal C}\left(k_{\text{max}},  \omega_{\text{max}}\right) + \tfrac{1}{\Delta k} \int_{0}^{k_{\text{max}}} \mathrm{d} k \; \log g_{\bar{H}}(k) \bigg]\,.
\end{align}
As argued in section~\ref{subsec:one cont Fourier}, even though the exact form of~$g_{\bar{H}}(k)$ in~\eqref{eq:kpz-delta-i-result} is complicated in this example, it suffices to consider its Taylor expansion around $k = k_0$ and $\bar{H} = \bar{H}_{\text{c}}(\ell \to \infty)$. The result of this is shown for an example value of $\delta \bar{H} = \bar{H}_{\text{c}}(\ell \to \infty) - \bar{H} = 0.1$ in Fig.~\ref{fig:g-approx-kpz} and reads $g_{\bar{H}}(k) = A \delta \bar{H} + \tilde{A} (\delta \bar{H})^2 + B (\delta k)^2 + \tilde{B} (\delta k) (\delta \bar{H})$ up to second order. Following the calculation of~\ref{subsec:one cont Fourier} shows that the mixed term due to $\tilde{B} \neq 0$ does not change the result, and we still end up with $\lambda = 4/3$ and $\theta = 2/3$ in the present example since $n = 1$.

To further cement the predictions of the Gaussian fluctuation method, we now turn to a different set of hydrodynamic theories -- the macroscopic fluctuation theory. We will recall the results of \cite{shpielberg-nemoto-caetano:2018}, and recast them in the language of the Gaussian fluctuation method.

\section{Macroscopic Fluctuation Theory\label{sec:MFT}}

The MFT is a theoretical framework developed to describe and analyze fluctuations in diffusive and interacting nonequilibrium systems. The central idea behind the MFT is to provide a hydrodynamic description of path probabilities, which in turn allow to calculate relevant hydrodynamic observables. 

The MFT is built on the assumption that the large-scale behavior of fluctuating fields (such as density or current) in nonequilibrium systems can be described by a  hydrodynamic Langevin equation. In simple cases, one can derive the hydrodynamic Langevin equation from lattice-gas models through coarse-graining procedures~\cite{doi:1976,peliti:1985,lefevre-biroli:2007,baek2019finite}.
The theory has been validated extensively through comparisons with microscopic simulations and exact results of lattice models, such as the Symmetric Simple Exclusion Process and other interacting particle systems. These comparisons show that MFT correctly predicts not only typical behavior but also rare and large fluctuations, as well as DPTs in the space-time evolution of these systems. In particular, the MFT has been used to predict single file and first passage time statistics~\cite{krapivsky2014large,grabsch2024joint,kumar2024arrhenius,kumar2024emerging,agranov2018narrow}, nonequilibrium correlations and fluctuation induced forces~\cite{garrido2022correlations,grabsch2024particle,aminov2015fluctuation}, void formations \cite{krapivsky2012void} to name but a few. Moreover, it has revealed a rich set of DPTs~\cite{shpielberg2017geometrical,shpielberg2017numerical}. 

The setup typically consists of a set of interacting particles in a system of size $L$. Particles can only enter or leave the system through coupling to particle reservoirs.  For the sake of simplicity, we will consider here a periodic system, where particles are both locally and globally conserved. In the hydrodynamic limit, we focus on the particle density $\rho(x,t)$, which satisfies the continuity equation $\partial_t \rho = -\partial_x j$. Here, $x \in [0,1]$ is the rescaled position, which is assumed to belong to a periodic system, and $t\in [0,\mathcal{T}] $ is the diffusively rescaled time. The 
current density~$j$ follows 
\begin{equation}
    j = J(\rho) + \sqrt{\chi/L} \,\xi(x,t)\,,
\end{equation}
where $J(\rho) = -D\partial_x \rho + \chi F$ corresponds to Fick's law, driving particles against the density gradient with diffusivity $D=D(\rho
)$, and in the direction of an externally applied force $F$ with mobility $\chi = \chi(\rho)$. The density evolution follows the general form of~$\eqref{eq:Langevin equation}$. 

It can be understood that the particles' dynamics is captured by the diffusivity and mobility alone, highlighting the elegance of the MFT. We now focus on a known case where a DPT can be observed: The current fluctuations of the weakly asymmetric exclusion process (WASEP) on a ring:  
The WASEP describes a lattice-gas model where particles can move on a one dimensional lattice. Importantly, particles can only hop to vacant lattice sites, with rate $\exp \left\{\pm F/2L \right\}$, where the plus (minus) sign implies right (left) jumps. 
In the language of the MFT, $D(\rho)=1$, and $\chi=\rho(1-\rho)$.

Here, we are interested in the probability $\mathcal{P}(Q/L \mathcal{T} = j_0 )$ of observing an atypical charge flux $Q$ over the (hydrodynamic) time $\mathcal{T}$. To find $\mathcal{P}( j_0 )$, one needs to consider the MFT's path probability, conditioned to satisfy $\intop^1 _0 \mathrm{d} x \intop^\mathcal{T} _0 \mathrm{d} t \, j(x,t) = j_0 \mathcal{T}$, which plays the part of~\eqref{eq:z-condition-gen}. The optimal fluctuation, minimizing the action~\eqref{eq:action-func}, is given by $j(x,t)=j_0$ and $\rho(x,t)=\rho_0$, where $\rho_0$ is the mean particle density in the system. This leads to $I_0(j_0) =  \mathcal{T} (j_0-2\chi_0 F)^2 / (4\chi_0)$, with the small parameter  $\varepsilon = 1/L$ accounting for the system size $L$, and~$\chi_0 = \chi (\rho_0)$.  

Adding perturbations to the optimal fluctuation allows to test the validity of this solution. Taking $\rho \rightarrow \rho_0 + \delta \rho / \sqrt{L}$, $j \rightarrow j_0 + \delta j / \sqrt{L}$, we can expand the action~\eqref{eq:action-func} to second order in the perturbations.  Again, we perform the expansion in the Fourier representation, which, to satisfy the current condition as well as the continuity equation, is given by
\begin{eqnarray}
    \delta \rho(x,t) &=& \sum_{\substack{k \in 2 \pi \mathbb{N}\\ \omega \in 2 \pi \mathbb{Z} / {\cal T}}} k a_{k,\omega} e^{i k x + i \omega t } + k a^* _{k,\omega} e^{-i k x - i \omega t }\,, \nonumber
    \\ 
    \delta j(x,t) &=& -\sum_{\substack{k \in 2 \pi \mathbb{N}\\ \omega \in 2 \pi \mathbb{Z} / {\cal T}}} \omega a_{k,\omega} e^{i k x + i \omega t }+ \omega a^* _{k,\omega} e^{-i k x - i \omega t }\,.
\end{eqnarray}
With this notation in mind, we find that
$\mathcal{P}(j_0)\simeq e^{-L I_0(j_0)}\int \prod_{k,\omega} \mathrm{d} a_{k,\omega}  \mathrm{d} a_{k,\omega}^* \, e^{-\mathcal{T} |a_{k,\omega}|^2 g_{j_0}(k,\omega)}$, where 
\begin{equation}
    g_{j_0}(k,\omega) = \mathfrak{a} \, \omega^2  +\mathfrak{b}  \, k^4 +
    \mathfrak{c} \, k^2 + \mathfrak{d} \, k \omega . 
\end{equation}
Above, we have defined the coefficients
\begin{align}
\begin{cases}
    \mathfrak{a} = \mathfrak{b} = 1/2\chi_0, 
    \quad 
    \mathfrak{d}= j_0  \chi_0 ' / \chi_0^2,  \\  
    \mathfrak{c} = F^2 -  j_0^2 \left[ (\chi_0 ')^2 - \frac{1}{4}\chi_0 \right]  / \chi_0^3\,. 
\end{cases}
\end{align}
Here, we consider the large macroscopic time $\mathcal{T}$ to be large, such that $\sum_\omega \rightarrow \mathcal{T}/ (2\pi) \cdot \int \mathrm{d} \omega$. With that in mind, we expect to obtain the CEs of case A, i.e.\ section~\ref{subsec:one cont Fourier}. To see that, we can sum up  the subleading contributions~\cite{appert-roland-derrida-lecomte-etal:2008}
\begin{eqnarray}
    \mathcal{P}(j_0) &\simeq& e^{-L I_0(j_0)} 
    e^{\mathcal{T} D \mathcal{F}(u)}, 
    \\ \nonumber 
    \mathcal{F}(u) &=& - \sum_{k} \sqrt{k^4 -8 u k^2 } - k^2 +4u, 
\end{eqnarray}
     where $u = (j^2_0-\chi^2 _0 F^2)\chi_0 '' / (16D\chi_0)$. 
The important frequency tuple can be seen to be  $(k_0,\omega_0) = (2\pi,0)  $. Then, we find that the dominating singular contribution in~$\delta I$ is given by 
$ I_\text{s} \propto -\mathcal{T}\sqrt{\delta u} / L$, where  $\delta u = u_{\text{c}}-u$ and $u_\text{c}= \pi^2 /2$. This implies that $n=1$ in~\eqref{eq:scaling n}. As predicted from the classification to case A, the analysis leads to $\lambda=4/3$, $\theta=2/3$ which were identified already in \cite{shpielberg-nemoto-caetano:2018,baek2018dynamical}. 

\section{Summary and discussion \label{sec:Summary} }

In this work, we have investigated the CEs associated with continuous DPTs in weak noise Langevin systems. Our focus was on the probability of observing atypical events, which we characterized using a large deviation framework. In the context of weak noise Langevin dynamics, the large deviation function is primarily determined by an optimal fluctuation. Close to a DPT, accounting for fluctuations about the optimal fluctuation leads to an anomalous correction to the large deviation function in the form of a singular scaling function~\eqref{eq:I singular scaling form}. By employing the Gaussian fluctuation method introduced here, we classified the CEs $\lambda,\theta$ governing the behavior of this singular scaling function. Previously, identifying the CEs in weak noise theories was possible mainly due to the identification of an order parameter, and developing a Landau theory, suggesting that only a limited set of problems could be handled. The Gaussian fluctuation method introduced here offers a broader approach to identifying the CEs $\lambda,\theta$.

We have demonstrated that the CEs $\lambda,\theta$ can be classified into three distinct universality classes, depending on the boundary conditions of the system, when considering a single fluctuating field in $1+1$ dimensions. Extending the analysis to higher dimensions or to systems with multiple fluctuating fields is conceptually straightforward. For higher spatial dimensions, the Fourier modes $k$ become vectors. The task reduces once again to count the number of required continuous Fourier modes which leads to more possibilities. One would need to perform higher dimension integrals, like \eqref{eq:two cont. integral}, to infer the CEs. For the case of additional fluctuating fields, one would express them using the Fourier decomposition as well, and obtain matrix-valued coefficient functions $g_z(k, \omega)$. Typically, one expects that at the critical point, it is only one of the 
eigenvalues of $g_z(k_0, \omega_0)$
that becomes zero, thus recovering our previous treatment.

However, it is important to note that numerical or experimental verification becomes increasingly challenging beyond $1+1$ dimensions.

In this work, we have demonstrated the Gaussian fluctuation method of classifying the CEs for two cases: DPTs in the short-time KPZ equation and diffusive systems within the framework of the MFT. One may ask whether the Gaussian fluctuation method could still be valid for theories with long-range interactions where the action may become non-local \cite{dandekar2023dynamical}, or to include memory effects where additional fluctuating fields may be present \cite{woillez2019activated}. For the case of additional fluctuating fields was already discussed, and the theory remains intact. For non-local actions, the theory still applies as well. However, providing an tractable example where one can still fully diagonalize the Gaussian fluctuations, i.e.\ find $g_z (k,\omega)$ explicitly, will be particularly challenging due to the non-locality of the action. Hence, demonstrating the Gaussian fluctuation method for such a system remains a future challenge. Finally, we do not expect the Gaussian fluctuation theory to hold for BKT phase transitions.

Although three universality classes of the CEs $\lambda,\theta$ were identified in this work, only two were demonstrated in physical systems. An intriguing future direction would be to explore a weak noise system where the third universality class can be realized. The third universality class requires taking the joint limit of large system size and observation time. This violates the diffusive scaling of the MFT, implying one should look elsewhere for a physical demonstration. Additionally, all examples considered in this work involve continuous DPTs of order $G_{\text{c}} = 2$, with $n = 1$ for the unstable frequency coefficient in~\eqref{eq: k0 omega0 f}. Finding and studying physical systems with other values of these parameters, perhaps even with vanishing denominators in~\eqref{eq:crit-expo-one-continous} or~\eqref{eq:crit-expo-two-continous}, would be interesting.

We point out that often weak noise Langevin systems have, in addition to the fluctuating field $\phi$, a conjugate field, stemming from a local conservation law. In fact, this is ever present in the MFT, where particle conservation is assumed throughout. Our analysis remains unaffected by the addition of local conservation laws, as can be observed from section~\ref{sec:MFT}.   

The CEs $\lambda,\theta$ obtained in this work differ from the textbook CEs $\alpha, \beta, \gamma, \delta, \nu$, which are associated with the specific heat, order parameter, susceptibility, and correlation length in thermodynamics~\cite{ma:2018,kardar:2007,cardy:1996,cardy:2012,stanley:1971}. In the following, we argue that $\alpha$ and $\nu$ are related to $\lambda$ and $\theta$. However, the other textbook exponents $\beta$, $\gamma$, and $\delta$, which are defined with respect to the order parameter, cannot be deduced solely from the analysis of $\lambda$ and $\theta$.

First, recall that $\alpha$ is related to the specific heat, which is the second derivative of the free energy density with respect to temperature. In our context, connecting this to the large deviation function $I$, we define the specific heat as $C(z) = \partial_{zz} I(z)$ with $C(z) \sim \left(\delta z\right)^{-\alpha}$ close to the DPT. From the singular scaling form of $I$ in~\eqref{eq:I singular scaling form}, we derive $\alpha = (\lambda - 2\theta)/\theta$, where we have imposed $\varepsilon = \delta z^{1/\theta}$. The Ginzburg criterion then ensures that $\alpha = 0$, consistent with the mean-field theory prediction.

Second, for the $1+1$ dimensional problem studied here, there are two correlation lengths: the spatial correlation length~$\xi_x$ and the temporal correlation time~$\xi_t$ associated with the dynamical exponent. We propose that near criticality, $I \sim (\xi_t \xi_x)^{-1} \sim \left( \delta z \right)^{-\nu d_\perp}$, where~$d_\perp$ is the effective number of continuous Fourier modes. Notice that for~$d_\perp = 0$, $\xi_i$ with $i \in \lbrace x,t \rbrace$ does not diverge.   Otherwise, we recover the scaling relation $2 - \alpha = \nu d_\perp$. Importantly, the value of $\nu$ may differ from the mean-field exponent $\nu = 1/2$, which is valid for the critical dimension $d = d_{\text{c}} = 4$ in thermodynamics. We note that depending on the continuous spatial/temporal degrees of freedom, the correlation scales $\xi_{x,t} $ either diverge with exponent $\delta z ^{-\nu}$ or remain finite.

Furthermore, recall that in thermodynamics, the textbook CEs are all connected with empirically measurable quantities. However, in the context of large deviations, it may be more feasible to determine the CEs $\lambda$ and $\theta$
directly through the finite-size corrections to the large deviation function $I$. See~\cite{shpielberg-nemoto-caetano:2018} for a demonstration of these CEs in a surprisingly small system. In contrast, in systems where the order parameter is non-trivial, it can be challenging to numerically verify the CEs, such as $\beta$, due to the complexities involved in accurately measuring the order parameter's behavior near criticality, as in~\cite{smith-kamenev-meerson:2018} where the asymmetry of a one-dimensional KPZ interface with respect to $x = 0$ is used as the order parameter for a mirror-symmetry breaking DPT. We also comment that while an experimental study of large deviations and dynamical phase transitions seem challenging, several works have recently made advances into observation of large deviations in general \cite{agranov2020airy}, and DPTs in particular \cite{kumar2024inferring}.

Analytically and numerically determining the shape of the singular scaling function and subleading terms would be another interesting future direction, cf.~\cite{stella-chechkin-teza:2023,balo-delamotte-rancon:2024}. In section~\ref{Sec:Infinite range Ising}, we have shown that to recover the full singular scaling function $\mathfrak{f}_{\text{s}}$, it is required to go beyond the Gaussian fluctuations. This should be the case for weak noise Langevin systems as well. 

Note that similarly to thermodynamic phase transitions, subleading singular contributions are present and can be observed from the calculations. Therefore, numerical values of the CEs $\lambda,\theta$ could be different than expected when $\varepsilon$ is not sufficiently small to differentiate between the competing singular terms. This hurdle is of course well known in the context of thermodynamic phase transitions, see the discussion in~\cite{cardy:2012}. 

It is important to note that numerically determining the CEs of a continuous DPT associated with a large deviation function in weak-noise Langevin systems via Monte Carlo simulations is challenging. Even away from a DPT, simulating rare, atypical events typically requires large simulation data sets to capture the constrained trajectory ensemble subject to the constraint~\eqref{eq:z-condition-gen} with sufficient statistical accuracy. This is due to the fact that the number of samples needed for a given relative accuracy scales with $1 / {\cal P}(z)$ for ${\cal P}(z) \ll 1$, unless specialized importance sampling schemes are employed~\cite{bucklew:2013,hartmann-le-doussal-majumdar-etal:2018,ebener-margazoglou-friedrich-etal:2019,perez-espigares-hurtado:2019,hartmann-krajenbrink-le-doussal:2020}. However, close to the critical point, the number of trajectories required again increases dramatically, because a higher accuracy in adhering to the constraint~\eqref{eq:z-condition-gen} is required. Therefore, one may expect that extracting the CEs is difficult in general. 

In this vein, numerically verifying the CEs predicted here could be considered a high precision test for different rare event samplers of weak noise Langevins systems. That is, if a code can accurately measure the CEs, it demonstrates the precision and efficiency of the numerical method to predict large deviations. 

As a final remark,  we have assumed throughout this work that the hydrodynamic theory encapsulates the range of possible phenomena, and furthermore that microscopic agents cannot affect its behavior, leading to independence in choosing microscopic cutoffs. While this is generally true, one can point out cases where these assumptions break down, e.g.\ condensation of particles on a microscopic length scale \cite{evans2005nonequilibrium}, shock waves in gases due to a moving piston \cite{zel2002physics}, and effects of a local battery drive \cite{sadhu2011long,kado-sasa-2024}. It would be interesting to develop a rigorous criterion, revealing when hydrodynamic theories are self-consistent, such that microscopic cutoffs cannot alter the macroscopic behavior.

\paragraph*{Acknowledgments.} The authors thank Ori Hirschberg for proposing the CW model as a pedagogical example, and gratefully acknowledge Alexander Hartmann for beneficial discussions on numerical verification of the CEs.   

\appendix 

\section{Full singular scaling function for the Curie--Weiss model}
\label{sec:appendix-cw}

For completeness, we evaluate the full singular scaling function $\mathfrak{f}_{\text{s}}$ of the CW model with Hamiltonian~\eqref{eq:C-W model Ham} here. Taylor expanding $g_z(m)$ as defined in~\eqref{eq:g-def-cw-model} around $m=0$, we obtain
\begin{equation}
    -N g_z(m) = N \log 2 - \frac{1}{2} z \delta z N m^2 - \frac{z^4}{12} N m^4 + O\left(z^6 N m^6\right)\,.
\end{equation}
Close to the transition such that $\delta z \downarrow 0$, we consider $\mu = \delta z N^{\theta} = \text{const.}$ with finite $\mu > 0$ and $\theta$ to be determined. It is useful to redefine $y= N^{\frac{1-\theta}{2}} m$. We thus find  
\begin{align}
    {\cal Z}_N  = & \sqrt{ \frac{N^\theta z}{ 2\pi }} e^{N \log 2}  \int_{-\infty}^\infty \mathrm{d} y  \, e^{-\frac{1}{2} \mu z y^2 - \frac{1}{12} N^{2 \theta - 1} z^4 y^4 +O \left(N^{3 \theta - 2}\right) }\nonumber\\
     \simeq & \sqrt{ \frac{N^\theta z}{ 2\pi} } e^{N \log 2}  \int_{-\infty}^\infty \mathrm{d} y  \, e^{-\frac{1}{2} \mu y^2 - \frac{1}{12} y^4 } \text{ for } \theta = \frac{1}{2}\,,
\end{align}
as $N \to \infty$ and $\delta z \downarrow 0$ while $\mu = \text{const.}$
Here, we have determined the critical window by requiring that quartic fluctuations become comparable to the quadratic ones. With this, we find the full singular scaling function 
\begin{equation}
    \mathfrak{f}_N(z) = - \log 2  + \frac{1}{N} \mathfrak{f}_{\text{s}}\left( \mu = \delta z N^{1/2}\right), 
\end{equation}
where $\mathfrak{f}_{\text{s}}\left(\mu \right) = -\log \int_{-\infty}^\infty \mathrm{d} y  \, e^{-\tfrac{1}{2}\mu y^2 - y^4 / 12  }$, which has an explicit expression in terms of the Bessel function $K_{-1/4}$.

\bibliography{bib}

\end{document}